\documentclass[reprint,amsmath,amssymb,aps,prl,superscriptaddress]{revtex4-1}

\usepackage{times,color,amsmath}
\usepackage{subfigure,epsfig,graphicx,epstopdf}
\usepackage[colorlinks]{hyperref}
\hypersetup{linkcolor = {blue},	citecolor = {blue},	urlcolor = {blue}}
\usepackage{textcomp}

\newcommand{\affa}{\affiliation{Center for Integrated Quantum Information Technologies (IQIT), School of Physics and Astronomy and State Key Laboratory of Advanced Optical Communication Systems and Networks, Shanghai Jiao Tong University, Shanghai 200240, China}}
\newcommand{\affb}{\affiliation{CAS Center for Excellence and Synergetic Innovation Center in Quantum Information and Quantum Physics, University of Science and Technology of China, Hefei, Anhui 230026, China}}
\newcommand{\affc}{\affiliation{School of Mathematical and Physical Sciences, University of Technology Sydney, Ultimo, New South Wales 2007, Australia}}

\begin{document}

\title{Topologically Protecting Squeezed Light on a Photonic Chip}

\author{Ruo-Jing Ren} \affa \affb         \author{Yong-Heng Lu} \affa \affb        \author{Ze-Kun Jiang} \affa \affb		 
\author{Jun Gao} \affa \affb          		\author{Wen-Hao Zhou} \affa \affb        \author{Yao Wang} \affa \affb
\author{Zhi-Qiang Jiao} \affa \affb          \author{Xiao-Wei Wang} \affa \affb        \author{Alexander S. Solntsev} \affc
\author{Xian-Min Jin} \email{xianmin.jin@sjtu.edu.cn}  \affa \affb

\date{\today}

\begin{abstract}
Squeezed light is a critical resource in quantum sensing and information processing. Due to the inherently weak optical nonlinearity and limited interaction volume, considerable pump power is typically needed to obtain efficient interactions to generate squeezed light in bulk crystals. Integrated photonics offers an elegant way to increase the nonlinearity by confining light strictly inside the waveguide. For the construction of large-scale quantum systems performing many-photon operations, it is essential to integrate various functional modules on a chip. However, fabrication imperfections and transmission crosstalk may add unwanted diffraction and coupling to other photonic elements, reducing the quality of squeezing. Here, by introducing the topological phase, we experimentally demonstrate the topologically protected nonlinear process of spontaneous four-wave mixing enabling the generation of squeezed light on a silica chip. We measure the cross-correlations at different evolution distances for various topological sites and verify the non-classical features with high fidelity. The squeezing parameters are measured to certify the protection of cavity-free, strongly squeezed states. The demonstration of topological protection for squeezed light on a chip brings new opportunities for quantum integrated photonics, opening novel approaches for the design of advanced multi-photon circuits.
\end{abstract}
\maketitle

Over the last few decades, researchers have witnessed the emerging field of quantum information\cite{NielsenChuang}. Various advances have been achieved in a plethora of hardware platforms\cite{QC,QQ}. Photon, due to its fast spread and robustness against the thermal environment, is considered as a perfect information carrier for quantum information processing\cite{Photon1,Photon2}. Thus, quantum light sources, particularly indistinguishable correlated photon pairs, are kernel resource for quantum communication\cite{QCrypt1,QCrypt2} and quantum computation\cite{HOM,QComp,N00N,Molecules}. Besides these researches focusing on Fock-like states, squeezed light also serves as another fundamental resources for quantum information\cite{CV1,CV2,GBS,BECC}. Spontaneous parametric frequency conversion in nonlinear crystals is an indispensable approach in quantum optics to obtain the non-classical light\cite{SPDC,Lightrev}, as well as to generate multi-photon entangled states\cite{Ent1,Ent2,Ent3,Ent4}. 

In order to obtain a strong nonlinear interaction in a bulk crystal, a tight focusing condition is required, however, which would lead to a small Rayleigh length limiting the interaction volume. The pump light power consumption in the bulk crystals is always too large for scalability, and the pump power usually exceeds the threshold of the material, causing modulations to the micro-nano structure leading to undesired nonlinear effects. In the last few years, integrated photonics stands out by offering a compact solution\cite{InteP1,InteP2}, which greatly reduces such undesirable effects by confining the pump light inside the waveguide, thus both of the tight focusing condition and interaction zone are guaranteed at the same time. The progresses in this field have been demonstrated by developing waveguide-based quantum sources by femtosecond laser writing\cite{Spring2013}, UV-laser writing\cite{Spring2017}, and silicon photonics platforms\cite{Silicon2014,Silicon2020}, showing the availability and the high performances like brightness, purity and low propagation loss in a single chip. Most recent work has shown the potential of integrated source to encode information in both discrete and continuous variables\cite{Ren}.

To construct large-scale photonic quantum systems, high-quality building blocks should be integrated in a compact footprint with an immense complexity\cite{JWang2018}. Compared to the silicon photonics platform, femtosecond laser writing chips show the capability of fully 3D integration, which is particularly suitable for simulating 2D structures\cite{Fab,3DFab}, and have been used to carry out various quantum tasks\cite{2DQW,Fast_hitting}. Supposing an on-chip source module is embedded into a complex arrangement, due to the coupling effect, crosstalk is prone to occur between adjacent waveguides. Furthermore, the intensity to maintain strong squeezing will be greatly reduced during the evolution, since the pump light would diverge into the neighboring sites. These unwanted losses break the strong confinement and cause insufficient light-matter interaction to work as a quantum squeezer. The quantum features, for example, the cross correlation and the squeezing quality of the generated photon pairs will decay as time evolves. The key challenge is to protect the nonlinearity, the related interaction process and the quantumness of the generated photon pairs simultaneously.

\begin{figure*}
 \centering 
 \includegraphics[width=1.95\columnwidth]{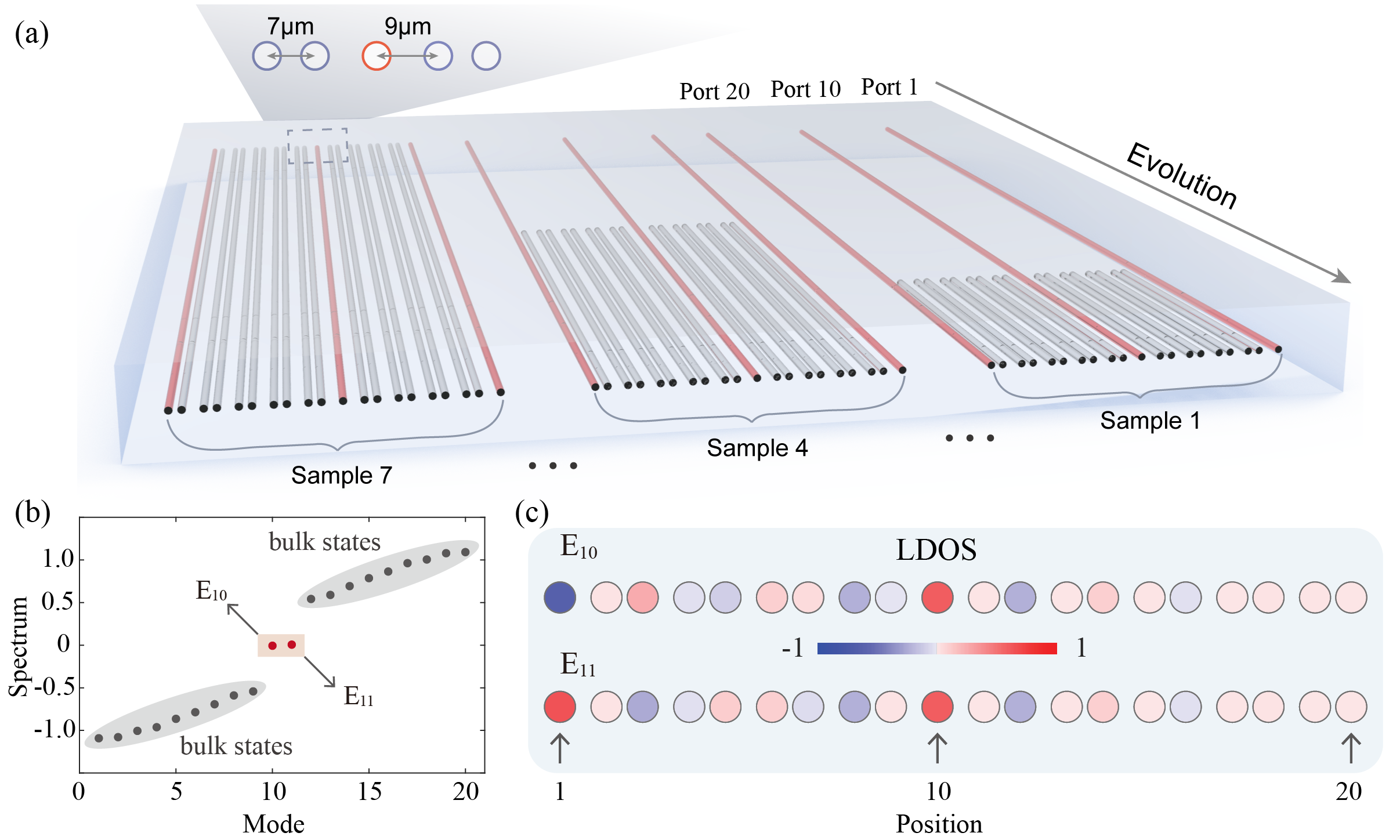}\\
 \caption{{\bf Schematic diagram of lattice with topological protection for squeezed light.} (a) The structure of the lattice with topological protection for squeezed light contain seven groups with evolution distances ranging from 5 mm to 35 mm. The short and long separation between adjacent waveguides is 7 \textmu m and 9 \textmu m respectively. (b) The spectrum of the lattice with topological protection for squeezed light. There are only two zero-energy modes $E_{10}$ and $E_{11}$ and two extended bulk bands. Modes $E_{10}$ and $E_{11}$ locate inside the band gap and decouple from the bulk bands, which promises the topologically protected fashion. (c) The eigenmode local density of states (LDOS) of the gapped mode $E_{10}$ (upper line) and $E_{11}$ (lower line) with spatial distribution. The edge states and interface-defect states are mainly localized at the $1^{st}$ site and the $10^{th}$ site of lattice both with high amplitude.}
 \label{FIG. 1.}
\end{figure*}

The topological phase allows the protection of physical fields against unavoidable disorder leading to recent demonstrations of topological protection of various nonlinear optical processes\cite{Leykam2015,NonlinearRev}, including photon-pair generation\cite{Redondo2018,Wang2019,Tschernig2021,Rechtsman2016,Wang1,Wang2}. The above, however, was only studied in the regime of low photon numbers, while strongly squeezed light remained out of scope despite the significant importance of squeezing in quantum optics. In this letter, we experimentally exhibit the topological protection of the squeezed light with dimerized-type chain resembling Su-Schrieffer-Heeger (SSH) lattices on a photonic chip. We observe robust localization phenomenon of the topological states at different wavelengths. We verify the topological protection of quantum resources by measuring the transport dynamics such as light filed distributions, the cross-correlations by switching incident pump light into different input ports. The squeezing parameters are particularly measured under different evolution distances. Our results demonstrate that the topological protection is robust to different wavelengths of non-classical states, and can help to construct quantum squeezers in an engineered structure against imperfections.

Our topologically protected lattices are inscribed on a 20 mm$\times$35 mm$\times$1 mm fused silica substrate by femtosecond laser direct writing. The writing laser has a wavelength of 513 nm, and a repetition rate of 1 MHz. Based on our previous work\cite{Ren}, the waveguide itself can function as a Spontaneous Four Wave Mixing (SFWM) source, following energy and momentum conservation, $2\hbar\omega_{p}=\hbar\omega_{s}+\hbar\omega_{i}$ and $2\hbar k_{p}=\hbar k_{s}+\hbar k_{i}$, where $k$ indicates the momenta of the field. The material absorbs two photons from the pump wave, and generates signal and idler photon pairs, where the birefringence induced phase matching condition is fulfilled by
\begin{equation}
\Delta k=\frac{2\left[n\left(\omega_{p}\right)+\Delta n\right] \omega_{p}}{c}-\frac{n\left(\omega_{s}\right) \omega_{s}}{c}-\frac{n\left(\omega_{i}\right) \omega_{i}}{c}=0.
\end{equation}
Here the birefringence $\Delta n$ dominates the phase matching condition.

The constructed topologically protected quantum light lattices, being composed of small and large spacing with adjacent waveguides, corresponding to the modulation of alternating weak ($J_{1}$) and strong ($J_{2}$) couplings, which can be described by the following Hamiltonian,
\begin{equation}
H=\sum_{n}\left(J_{1} a_{n}^{\dagger} b_{n}+J_{2} b_{n}^{\dagger} a_{n+1}\right) + h.c .
\end{equation}
The lattices possess two topologically protected channel: the edge-state channel and the interface-defect channel, as shown in Fig.1(a). Here, we set the separation distances $l_{1}$=7 \textmu m and $l_{2}$=9 \textmu m respectively. The dimerized-type chain resembling SSH model\cite{ssh} possesses and enables topological nontrivial phases with edge states described by the bulk-edge correspondence\cite{TPRev1,TPRev2}. The edge states can be regarded as the topological transition interface between trivial vacuum and nontrivial lattice structure. In addition, the interface-defect channel in the middle of the lattice acts as a topologically protected interface states\cite{Redondo2016} by interfacing two versions of the dimerization patterns with distinct Zak phases\cite{zak1,zak2}, which is supported by the existence of the topological phase transition between them.

\begin{figure}
 \centering
 \includegraphics[width=0.95\columnwidth]{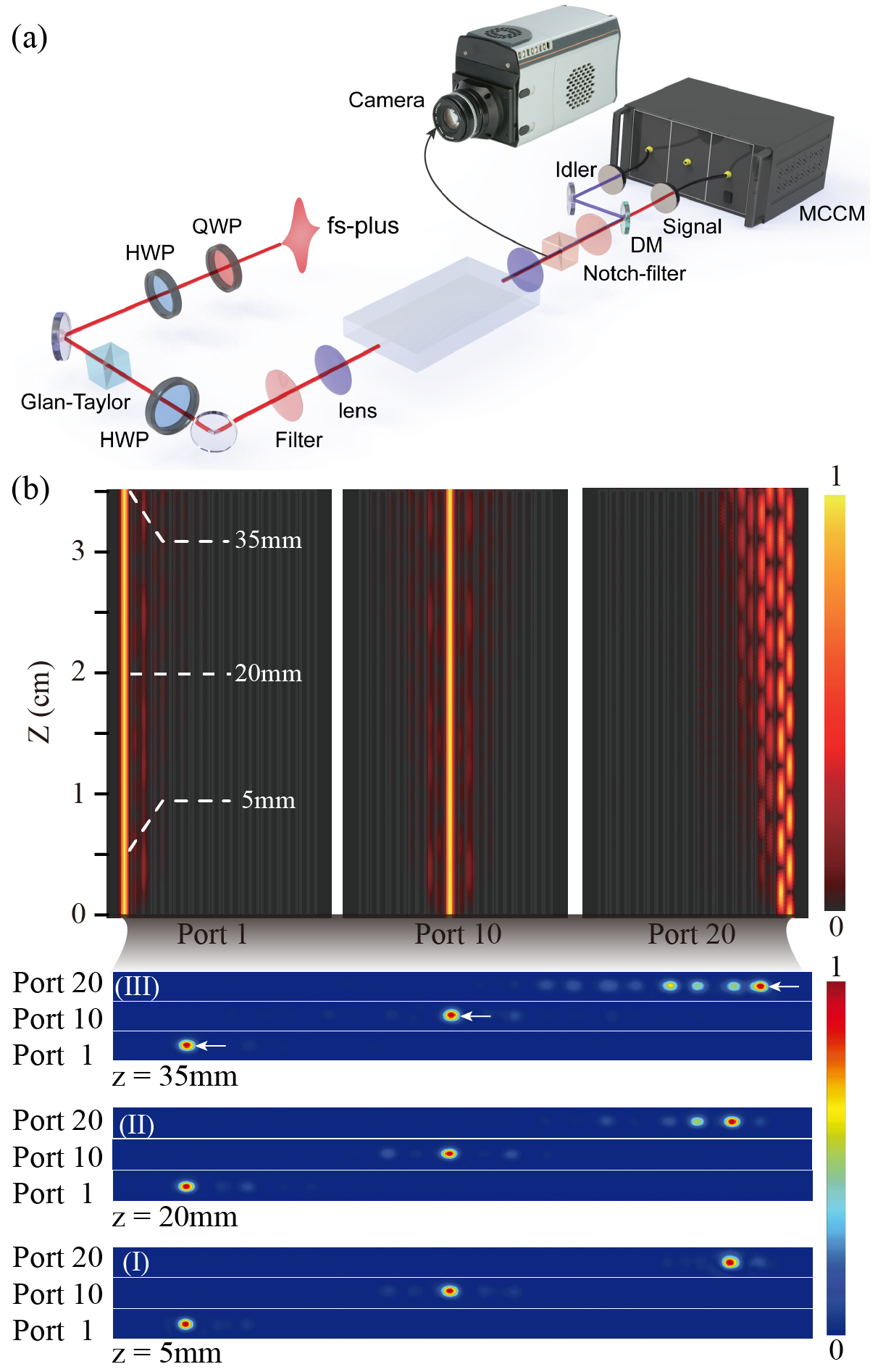}\\
 \caption{\textbf{Experimental setup and intensity distributions of the pump light.} (a) Experimental setup for generation, filtering and evolution measurement of the quantum light source. The input ports can be easily switched among different groups. DM: dichroic mirror, MCCM: multi-channel coincidence module. (b) Simulated intensity distributions of the pump light from port $1$, $10$ and port $20$. The evolution distance is marked in white color. (I)(II)(III) show the experimentally measured intensity distributions from three input channel ports at 5 mm, 20 mm and 35 mm evolution distances respectively. The protected states from port $1$ and $10$ always maintain localization as the traveling distance increases while the unprotected state from port $20$ diffuses over the lattice. The intensity distribution is normalized to its maximum.} 
 \label{FIG. 2.}
\end{figure}

We further illustrate the band diagram by characterizing the spectrum of the lattice, as shown in the Fig.1(b). The band diagram contains two extended bands separated by the band gap. Inside the band gap, there are two modes $E_{10}$ and $E_{11}$ pinning on the quasi zero energy level with decoupling from the bulk band, which reveals the existence of the topological gapped edge states. The spatial distribution can be characterized by the eigenmode local density of states (LDOS)\cite{Noh2018}, which can be defined by $D_{n}(E)=\sum_{m}\delta (E-E_{m})\psi_{n}^{m}$, where $E$ is the energy of the $m^{th}$ eigenstate $\psi (m)$ and $n$ is the site label. As shown in Fig.1(c), the LDOS of the two gapped modes indicates that the modal amplitude with maxima appears in sites $1^{st}$ and $10^{th}$ site, while the minima appears in other sites in both two modes. This shows the localization in the edge channel and interface-defect channel under the norm of the topologically protected zero-energy modes. In contrast, the low amplitude occupying at $20^{th}$ site shows that the edge channel in $20^{th}$ port is trivial with dominated bulk modes rendering photons diffuse into bulk of the lattice. Therefore, among the three input ports in the lattice with topological protection for squeezed light (depicted in Fig.1(b)), the edge-state channel port $1$ and the interface-defect channel port $10$ both are topologically protected, and the edge channel port $20$ is the trivial one. 

The experimental setup is schematically depicted in Fig.2(a), the mode-locked 780 nm femtosecond pump pulses (80 MHz) of vertical polarization (prepared by a combination of waveplates and Glan-Taylor polarizer) are injected and switched into three different channel ports, namely $1$,$10$ and $20$. The output light intensity distributions of the pump light are accumulated by a CCD camera, and the evolution patterns under different distances varying from 5 mm to 35 mm (see I, II, and III in Fig.2(b)) are also recorded. We experimentally verify the localization effect in the above 3 ports by comparing the measured pattern with theoretical simulation results in Fig.2(b). It is obvious that the pump light propagates locally as time evolves in the edge-state channel and the interface-defect channel (the topological ones), while the beam diffuses and gradually couples to the adjacent sites in the edge channel port $20$ (the trivial one). The localization effect of the above topological channel ports indicates valid protection of the pump intensity, and provides sufficient power for generating quantum source during nonlinear interaction process. The diffusion degrades the intensity as the propagation length increases, and the decreased pump intensity fails to achieve a high quality quantum source during the SFWM process. It can be deduced that the nonlinearity processes in port $1$ and port $10$ are well protected at the wavelength of 780 nm. The pump light is spatially protected and the good localization effect makes the subsequent nonlinear process more efficient.

\begin{figure*}
 \centering
 \includegraphics[width=1.95\columnwidth]{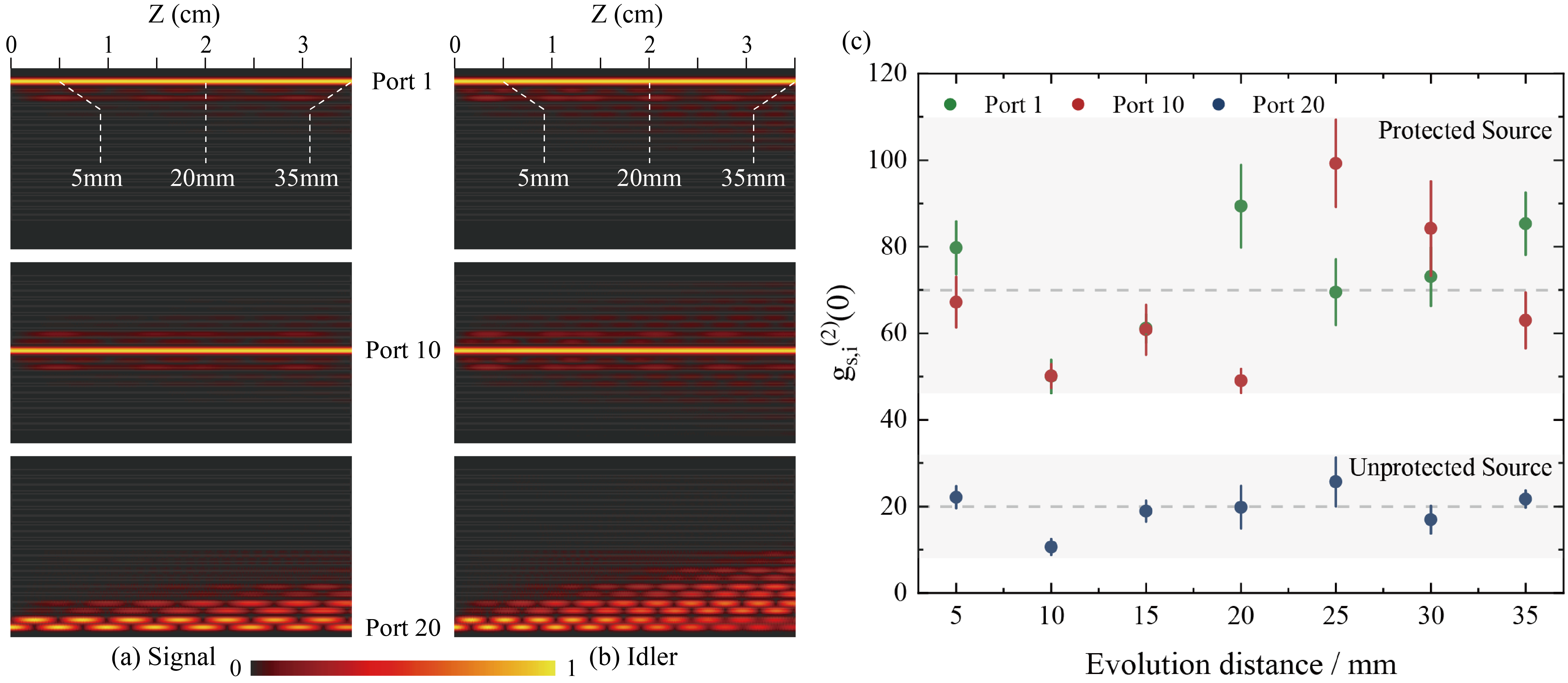}\\
 \caption{
  \textbf{Performance of topological protection for the generated photon pairs.} (a) The evolution result of the signal photon from different input ports. (b) The evolution result of the idler photon from different input ports. (c) The cross correlation $g_{si}^2(0)$ of different channel entrances $1$, $10$ and $20$ evolving from 5 mm to 35 mm with a step of 5 mm are measured. The cross correlation $g_{si}^2(0)$ of channel port $1$ and $10$ are depicted in green and red colors respectively with high values, which demonstrate the protection of the nonlinear process in the topological structures. The cross correlation $g_{si}^2(0)$ of channel port $20$ is depicted in blue color, and the values are about five times lower than the protected states from channel port $1$ and $10$.}
 \label{FIG. 3.}
\end{figure*}

\begin{figure}
 \centering
 \includegraphics[width=0.95\columnwidth]{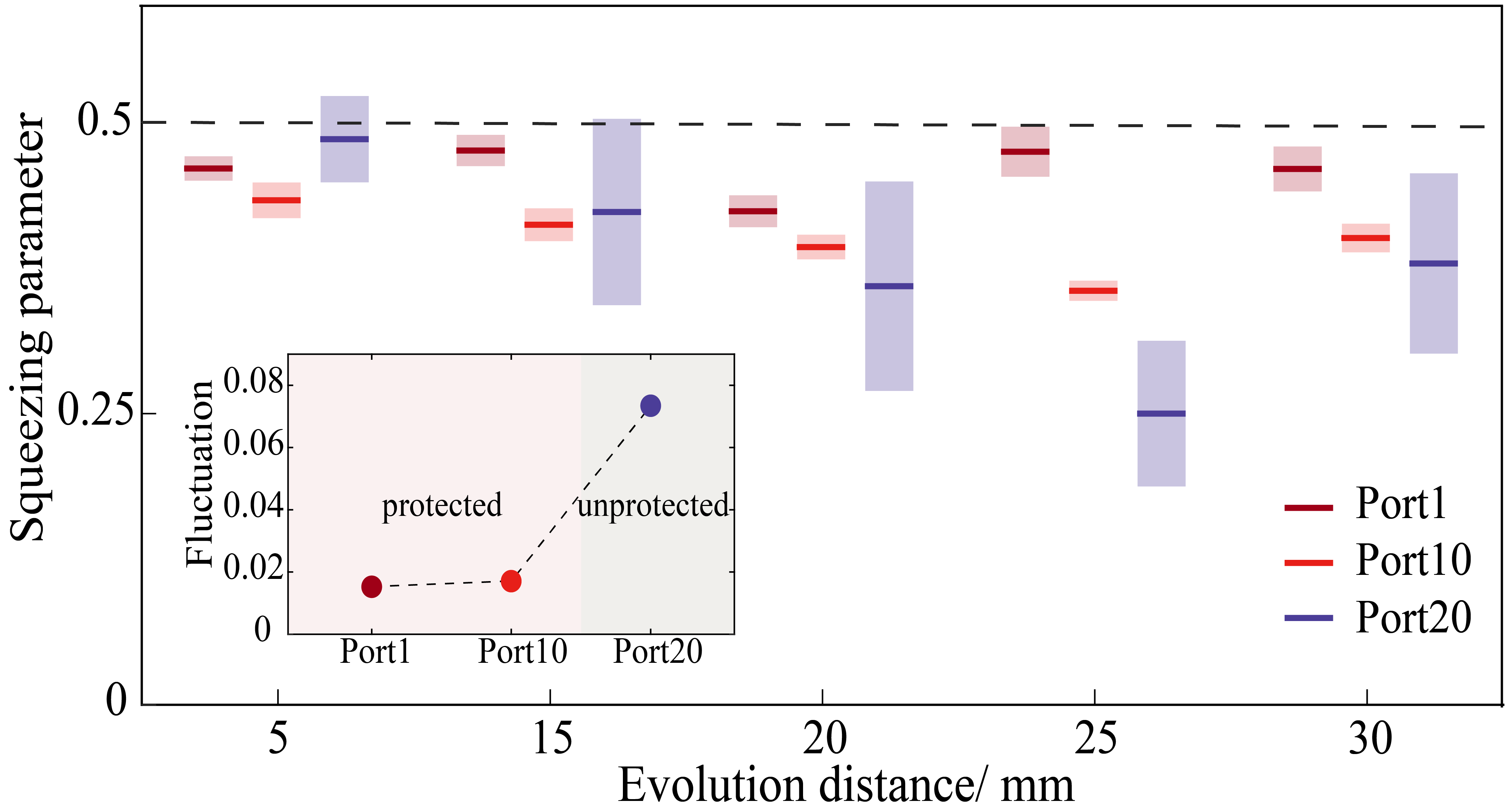}\\
 \caption{
  \textbf{Experimental verification of topological protection for the squeezing parameters.} The squeezing parameters of the channel port $1$, $10$ and $20$ are depicted in deep red, red and blue colors accordingly. The values of the squeezing parameters are more uniform and higher in two protected channel input ports compared with the unprotected channel port $20$. The variance between three topological channels are obviously different since the unprotected channel port $20$ fluctuates in a larger range. The inset shows the average fluctuation from each import. Compared to the results of the other two channel ports, the channel port $20$ is unprotected.}
 \label{FIG. 4.}
\end{figure}

After verifying the spatial protection effect of the pump light, we further test whether the on-chip generated correlated photons can still be well protected. We first calculate the light intensity distributions of the signal and idler photons, and the theoretical simulation results are presented in Fig.3(a) and (b) respectively. Even for light with different wavelengths, the localization effect still emerges in the topologically protected channels. To retrieve the signal and idler photons, we first filter out the residual pump light using both the polarization filter (Glan-Taylor polarizer) and the spectrum filter (notch filters centered as 780 nm). Then, the signal and idler photons are separated into two different spatial modes by a dichroic mirror, where the idler photons transmit while the signal photons are reflected. The photon pairs are coupled into single mode fibers, and then detected by avalanche photodiodes. All the coincidence counts are recorded by a homemade multi-channel coincidence module. The experiment layout can be found in Fig.2(a). To quantify the non-classical feature of the correlated photon pairs, we introduce the cross-correlation function as follows,
\begin{equation}
g_{s-i}^{(2)}(0)=\frac{\left\langle\hat{a}_{s}^{\dagger} \hat{a}_{i}^{\dagger} \hat{a}_{i} \hat{a}_{s}\right\rangle}{\left\langle\hat{a}_{s}^{\dagger} \hat{a}_{s}\right\rangle\left\langle\hat{a}_{i}^{\dagger} \hat{a}_{i}\right\rangle}
\end{equation}
A value higher than 2 of the cross-correlation function $g_{s-i}^{(2)}(0)$ is a strong evidence of quantum feature, for instance, the violation of Bell inequality requires $g_{s-i}^{(2)}(0)$ larger than 6\cite{stored}. 

The cross correlation $g_{s-i}^{(2)}(0)$ of the edge-state channel port $1$ vary from 50.03$\pm$3.85 to 89.34$\pm$9.53 with evolution distances from 5 mm to 35 mm, which is a strong evidence of non-classical correlation.The cross correlation $g_{s-i}^{(2)}(0)$ of the interface-defect channel port $10$ vary from 49.01$\pm$2.75 to 99.25$\pm$10.07, indicating the protection of SFWM photon pair generation process. Under the same transport length, in contrast, the cross correlation $g_{s-i}^{(2)}(0)$ of the unprotected edge channel port $20$ only vary from 10.58$\pm$1.82 to 25.65$\pm$5.6. As can be seen from the Fig.3(c), the topologically protected structure can provide nearly five times higher cross correlation than that in the unprotected states. If the evolution length continues to increase, the loss and decoherence will become more serious, and these eventually turn into the critical obstacles against protecting the nonlinear photon generation process. 

Due to the strong confinement of the pump light, our source can function as squeezed light source\cite{SV} in the high power regime. The threshold of fused silica is very robust, thus, combined with the engineered topological structures, both weak pump and strong pump regimes can be protected. In the weak pump regime, the discrete variables such as heralded photon sources are protected, while in the strong pump regime, the high-order terms of the nonlinear process and the squeezing parameters are dominant. We further explore the squeezing parameters among three different topological structures. The squeezing parameter $\lambda$ can be calculated by measuring the auto-correlation function $g_{H}^{(2)}(0)$ and the heralding efficiency $\eta_{H}$ using the following formula,
\begin{equation}
\lambda^{2}=g_{H}^{(2)}(0) \frac{\eta_{H}}{2\left(1-\left(1-\eta_{H}\right)^{2}\right)}.
\end{equation}
As shown in Fig.4, the squeezing parameters of the three channel ports are depicted in different colors under different evolution distances, where two distinct features can be observed. In the long evolution distance regimes, the squeezing parameters of the topologically protected channels are larger than the unprotected channels. In addition, the measured fluctuations using Possionian statistics in port $20$ are greatly influenced by the beam diffusion. These results indicate that the squeezed states can be well protected in the topological structure.


In conclusion, we have reported the topological protection of on-chip SFWM and the generated squeezed light. By introducing a topological phase with a dimerized-type chain resembling SSH lattices on a silica photonic chip, we have observed the localization and strong confinement of the pump light in the topologically protected channels. The protected pump light field fulfils the tight focusing condition to allow the waveguide to function as a high-quality quantum squeezer. We have demonstrated that the protection applies to different wavelengths, impacting the cross-correlations and the squeezing parameters. It showcases a robust generation of quantum resources for future practical quantum information tasks, such as Gaussian boson sampling applications and bosonic error correction codes. We have verified the validity of the topological protection of on-chip squeezers, which may play an essential role in photonic quantum information processing\cite{Schip}.\\

\begin{acknowledgments}

The authors thank Jian-Wei Pan for helpful discussions. This research is supported by National Key R\&D Program of China (2019YFA0308700, 2019YFA0706302 and 2017YFA0303700), National Natural Science Foundation of China (NSFC) (11904229, 61734005, 11761141014, 11690033), Science and Technology Commission of Shanghai Municipality (STCSM) (20JC1416300, 2019SHZDZX01), Shanghai Municipal Education Commission (SMEC)(2017-01-07-00-02-E00049). X.-M.J. acknowledges additional support from a Shanghai talent program. X.-M.J. acknowledges support from the National Young 1000 Talents Plan and support from Zhiyuan Innovative Research Center of Shanghai Jiao Tong University. A. S. S. Acknowledges support from Australian Research Council (DE180100070) and University of Technology Sydney Seed Fund.
	
\end{acknowledgments}

\end{document}